\documentclass[author-year, 5p]{elsarticle} %review=doublespace preprint=single 5p=2 column
\usepackage[hyphens]{url}
\usepackage{lineno} % add 
\biboptions{sort&compress} % For natbib
\usepackage{graphicx}
\usepackage{booktabs} % book-quality tables
\makeatletter
\def\ps@pprintTitle{%
 \let\@oddhead\@empty
 \let\@evenhead\@empty
 \def\@oddfoot{\it \hfill\today}%
 \let\@evenfoot\@oddfoot}
\makeatother

\usepackage[T1]{fontenc}
\usepackage{lmodern}
\usepackage{amssymb,amsmath}
\IfFileExists{microtype.sty}{\usepackage{microtype}}{}
\usepackage{longtable}
\usepackage{graphicx}
\makeatletter
\def\maxwidth{\ifdim\Gin@nat@width>\linewidth\linewidth
\else\Gin@nat@width\fi}
\makeatother
\let\Oldincludegraphics\includegraphics
\renewcommand{\includegraphics}[1]{\Oldincludegraphics[width=\maxwidth]{#1}}
  \usepackage[unicode=true]{hyperref}
\hypersetup{breaklinks=true,
            bookmarks=true,
            pdfauthor={},
            pdftitle={Early warning signals: The charted and uncharted territories},
            colorlinks=true,
            urlcolor=blue,
            linkcolor=magenta,
            pdfborder={0 0 0}}
\begin{document}
\begin{frontmatter}
  \title{Early warning signals: The charted and uncharted territories}
  \author[cstar]{Carl Boettiger\corref{cor1}}
  \ead{cboettig@gmail.com}
  \author[esp]{Noam Ross\corref{cor1}}
  \author[esp]{Alan Hastings}
  \cortext[cor1]{authors contributed equally}
  \address[cstar]{Center for Stock Assessment Research, Department of Applied Math and Statistics, University of California, Mail Stop SOE-2, Santa Cruz, CA 95064, USA}
  \address[esp]{Department of Environmental Science and Policy, University of California Davis, 1 Shields Avenue, Davis, CA 95616 USA}

  \begin{abstract}

The realization that complex systems such as ecological communities can collapse or shift regimes suddenly and without rapid external forcing poses a serious challenge to our understanding and management of the natural world.  The potential to identify early warning signals that would allow researchers and managers to predict such events before they happen has therefore been an invaluable discovery that offers a way forward in spite of such seemingly unpredictable behavior.  Research into early warning signals has demonstrated that it is possible to define and detect such early warning signals in advance of a transition in certain contexts.  Here we describe the pattern emerging as research continues to explore just how far we can generalize these results.  A core of examples emerges that shares three properties: the phenomenon of rapid regime shifts,  a pattern of 'critical slowing down' that can be used to detect the approaching shift, and a mechanism of bifurcation driving the sudden change.  As research has expanded beyond these core examples, it is becoming clear that not all systems that show regime shifts exhibit critical slowing down, or vice versa.  Even when systems exhibit critical slowing down, statistical detection is a challenge.  We review the literature that explores these edge cases and highlight the need for (a) new early warning behaviors that can be used in cases where rapid shifts do not exhibit critical slowing down, (b) the development of methods to identify which behavior might be an appropriate signal when encountering a novel system; bearing in mind that a positive indication for some systems is a negative indication in others, and (c) statistical methods that can distinguish between signatures of early warning behaviors and noise.
  \end{abstract}

 \end{frontmatter}

\section{Introduction}

Many natural systems exhibit regime shifts - rapid changes in the state
and conditions of system behavior. Examples of such shifts include lake
eutrophication (Carpenter et al. 1999), algal overgrowth of coral
systems (Mumby et al. 2007), fishery collapse (Jackson et al. 2001),
desertification of grasslands (Kéfi et al. 2007), and rapid changes in
climate (Dakos et al. 2008, Lenton et al. 2009). Such dramatic shifts
have the potential to impact ecosystem health and human well-being.
Thus, it is important to develop strategies for adaptation, mitigation,
and avoidance of such shifts.

The idea that complex systems such as ecosystems could change suddenly
and without warning goes back to the 1960s (Lewontin 1969, Holling 1973,
May 1977). Such early work revealed that even simple models with the
appropriate nonlinearities were capable of unpredictable behavior. The
only way to predict the transition was to have the right model -- and
that meant having already had the chance to observe the transition. One
cogent early example (Ludwig et al. 1978) demonstrated how knowledge of
the forms and time scales of interactions among insects, birds, and
trees could lead to a qualitative model that essentially predicted the
possibility of regime shifts.

Management of systems that could potentially undergo shifts requires
balancing the costs of adaptation, mitigation, or avoidance against the
costs of the shift itself. Avoidance depends on an ability to predict
regime shifts in advance, or depending on the time scale of response and
response of the system, on the ability to recognize a shift as it is
occurring. Adaptation and mitigation might require an ability to predict
a shift in advance if the time scale of implementation is long relative
to the rate at which damages occur.

An important component of this management challenge is the development
of early warning signals (EWS) of impending rapid regime shifts
(Scheffer et al. 2009). Since regime shifts occur in a variety of
systems, and underlying mechanisms for the shifts are not always known,
the development of generic signals applicable to a variety of systems
would be particularly valuable. This naturally leads to the questions of
when such generic signals would be valuable tools versus the need to
develop system-specific approaches in all cases.

Foundational research in EWS identified certain patterns that may
forecast a sudden transition in a wide variety of systems (Scheffer et
al. 2009). Most extensively researched is the phenomenon of critical
slowing down (CSD), which is manifested as a pattern of increasing
variance or autocorrelation of a system. Subsequent work has begun to
identify a growing library of cases in which these indicators are not
present before a transition (Schreiber and Rittenhouse 2004, Schreiber
and Rudolf 2008, Hastings and Wysham 2010, Bel et al. 2012), or are
observed in the absence of any transition (Kéfi et al. 2012). These
examples are distinct from the more well-known case of statistical error
-- such as a signal that is present but too weak to detect due to
insufficient available data (see Dakos et al. (2008); Scheffer et al.
(2009) and Perretti and Munch (2012)). Instead, such work moves into new
territory where different underlying mechanisms have lead to starkly
different patterns. Determining which underlying mechanisms are present
is a substantial empirical and theoretical challenge. When does critical
slowing down correspond to the assumptions made?

Here we review a variety of mechanisms that may lead to rapid (or
``catastrophic'') regime shifts in ecological systems, as well as
mechanisms that generate early warning signals. We focus on CSD and its
manifestations as they are the most commonly studied warning signals. We
illustrate that not all rapid shifts exhibit CSD, and not all
observations of CSD involve rapid shifts. Thus the issue of determining
EWS is really two-fold: first, to identify classes of systems where the
warning signal is expected and conversely systems that may undergo
shifts without such signals, and second, to determine appropriate
statistical tools to detect the warning signal. In this paper we review
both aspects of the overall question.

\begin{figure}[htbp]
\onecolumn
\begin{longtable}[l]{@{}ll@{}}
\hline\noalign{\medskip}
\begin{minipage}[t]{0.10\columnwidth}\raggedright
\textbf{Critical slowing down (CSD)}
\end{minipage} & \begin{minipage}[t]{0.35\columnwidth}\raggedright
A system's slowing response to perturbations as it's dominant eigenvalue
approaches zero, often expressed in greater variance, autocorrelation,
and return time. CSD is one possible EWS.
\end{minipage}
\\\noalign{\medskip}
\begin{minipage}[t]{0.10\columnwidth}\raggedright
\textbf{Early warning signals (EWS)}
\end{minipage} & \begin{minipage}[t]{0.35\columnwidth}\raggedright
A general term for dynamic patterns in system behavior that precede
regime shifts. Though CSD phenomena are among the best studied EWS, some
shifts will require alternative signals; Figure 1.
\end{minipage}
\\\noalign{\smallskip}
\hline
\end{longtable}
\twocolumn
\textbf{Definitions}: In this paper we refer to two closely related but
different phenomena
\end{figure}

\section{Relationships between Critical Slowing Down, Bifurcations, and
Regime Shifts}

CSD has been studied extensively in theoretical (Wissel 1984, Gandhi et
al. 1998, Carpenter and Brock 2006, Hastings and Wysham 2010, Dakos et
al. 2011a, Lade and Gross 2012, Boettiger and Hastings 2012a) and
empirical contexts (Drake and Griffen 2010, Carpenter et al. 2011,
Veraart et al. 2012, Dai et al. 2012, Wang et al. 2012) as a potential
EWS for regime shifts. CSD occurs as a system's dominant eigenvalue
approaches zero due to a changing (possibly deteriorating) environment.
As the eigenvalue approaches zero, the system's response to small
perturbations slows. This change in dynamic properties of a system can
be expressed in greater variance, autocorrelation, and return time of
observed state variables.

In Figure 1, we illustrate the domains of overlap between three distinct
phenomena. The first, \emph{rapid regime shifts}, are abrupt changes in
system behavior. The second, \emph{bifurcations}, are qualitative
changes in system behavior due to the passing of a threshold in
underlying parameters or conditions. Where these two overlap, we
sometimes call the phenomenon a ``catastrophic bifurcation.'' Finally,
\emph{critical slowing down} is the observed behavior of slow system
response to perturbation. The labels in italics describe examples of
phenomena that fall into these various domains. Below we describe cases
that fall into each of these regions.

\begin{figure*}[htbp]
\centering
\includegraphics{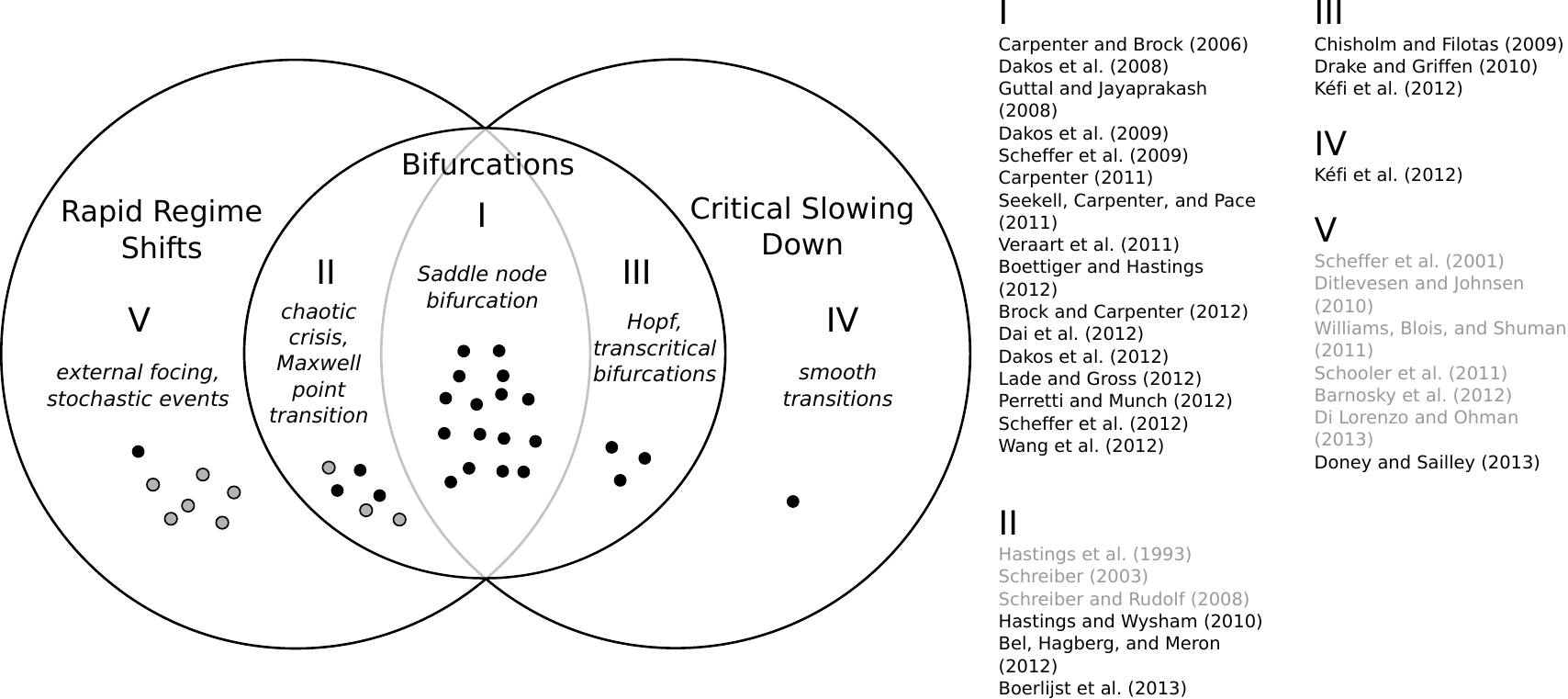}
\caption{Venn diagram representing the intersecting domains of rapid
regime shifts, bifurcations, and critical slowing down. Labels in italic
are example phenomena that occur in each domain. Roman numerals and
indicate example literature (right) exploring each domain, and also
refer to sections below describing those domains. Each dot represents a
study in the domain. Studies and dots in grey represent literature not
explicitly testing EWS, but which demonstrate phenomena related to EWS.
The center domain (I) where all three phenomena intersect, is the most
extensively researched domain of the EWS field. Literature outside this
charted region does not yet provide the needed EWS, but hints where
existing signals based on CSD may be insufficient or misleading.}
\end{figure*}

\subsection{Catastrophic Bifurcations Preceded by CSD (I)}

Much of the (most visible) recent research in EWS has focused on the
center of the diagram, where all three concepts intersect. The warning
signal patterns postulated, such as increasing variance and coefficient
of variation, (Carpenter and Brock 2006), increasing autocorrelation
(Dakos et al. 2008), increasing skewness (Guttal and Jayaprakash 2008a)
can all be directly derived from the changing eigenvalue in a saddle
node (also called fold) bifurcation. Consequently, experimental
evaluations of warning signals have largely focused on this situation as
well. CSD has frequently been studied in the context of models
exhibiting saddle-node bifurcations.

Dai et al. (2012) studied yeast cell growth in a microcosm and
demonstrated that an Allee effect created a saddle-node bifurcation in
the system. When the cell density was reduced to levels near the
bifurcation point, a decrease in recovery time (increase in variance and
autocorrelation over time) was observed. Veraart et al. (2012) studied a
system of cyanobacteria where models suggest a saddle-node bifurcation
driven by light inhibition. They also found increases in autocorrelation
and decreased recovery rates as the system approached the bifurcation.
These important experiments are among the best demonstrations that
saddle-node bifurcation dynamics really occur in natural systems, and
can be accompanied by reliable detection of EWS, at least when
sufficient data sampling, replicates, and controls are available.

Carpenter et al. (2011) provide a larger-scale example in which a lake
ecosystem is manipulated towards a sudden transition through the
introduction of a predator, while a neighboring experimental lake
provides a control. In this and similar lake systems, bifurcation is
thought to be driven in part by trophic interactions where adult fish
prey on the competitors of their juveniles (Carpenter and Kitchell 1996,
Walters and Kitchell 2001, Carpenter et al. 2008) which leads to a
saddle-node bifurcation. While the underlying dynamics of a whole lake
ecosystem are less tractable than the laboratory controlled chemoststats
of microorganisms, the system is understood well enough to anticipate
that a sudden transition can be induced under the intended manipulation.
Like the laboratory examples, this helps eliminate the options outside
the circle ``bifurcations,'' in Figure 1. The observed warning signals
then place it in the center of the diagram.

These studies have provided valuable demonstrations of the potential to
find early warning signals of sudden transitions. However, this
literature has begun to enumerate examples of similar transitions in
which no such signal is present.

\subsection{Catastrophic Bifurcations \emph{not} Preceded by CSD (II)}

Saddle nodes are only one of a variety of bifurcations, which can cause
rapid changes in system dynamics. Other bifurcations can cause long-term
changes in system dynamics without a gradual pass through a state with
zero eigenvalue, and therefore, not exhibit CSD. Many of these examples
can in fact show patterns in typical early warning indicator variables
such as variance or autocorrelation that are completely opposite to the
patterns seen in the saddle-node case. Several of these examples are
found outside the EWS literature, indicating a need to expand the range
of systems studied for EWS.

These are some of the most problematic cases. They represent disruptive
but potentially avoidable events, but would not be detected by using CSD
as an EWS. These cases include bifurcations in continuous time
(Schreiber and Rudolf 2008) and discrete time (Schreiber 2003),
explicitly spatial (Bel et al. 2012) and non-spatial, chaotic (Schreiber
2003, Hastings and Wysham 2010) and non-chaotic (Schreiber and Rudolf
2008, Hastings and Wysham 2010, Bel et al. 2012) examples. Before
warning signals can be reliably applied to novel systems, research must
provide a way to discern if the dynamics correspond to the better
understood warning signals of the saddle-node case or the more complex
patterns such as the examples discussed here.

One class of bifurcations in which we would not expect to see CSD prior
to regime shift are sometimes known as \emph{crises}. Crises are sudden
changes in the dynamics of chaotic attractors that occur in response to
small changes in parameters (Grebogi et al. 1983). Chaotic attractors
are features of many ecological models (Hastings et al. 1993), and
chaotic behavior has been shown in some ecological systems (Costantino
et al. 1997).

Hastings and Wysham (2010) examined a continuous model of a stochastic
three-species food chain where all species migrate between six patches.
When environmental stochasticity (represented as random variation in the
carrying capacity) is low, all species coexist in a chaotic but stable
attractor. A small increase in environmental stochasticity, though,
causes extinction of the top predator and rapid shift to a non-chaotic
cycle. Despite an increase in environmental variability, neither the
variance nor skew of the populations of any species change as the system
approaches this bifurcation.

Another example of a chaotic crisis can be found in a simple
discrete-time model where a population is subject to strong density
dependence (an Allee effect) and harvested by predators with a Type II
(saturating) functional response (Schreiber 2003). This case is
illustrated in Figure 2. When prey have high growth rates, the system
has chaotic dynamics. Small increases in the predation intensity cause a
bifurcation with chaotic but persistent prey populations to prey
extinction. As predation intensity increases towards this threshold, the
population exhibits \emph{decreasing} variance.

\begin{figure}[htbp]
\centering
\includegraphics{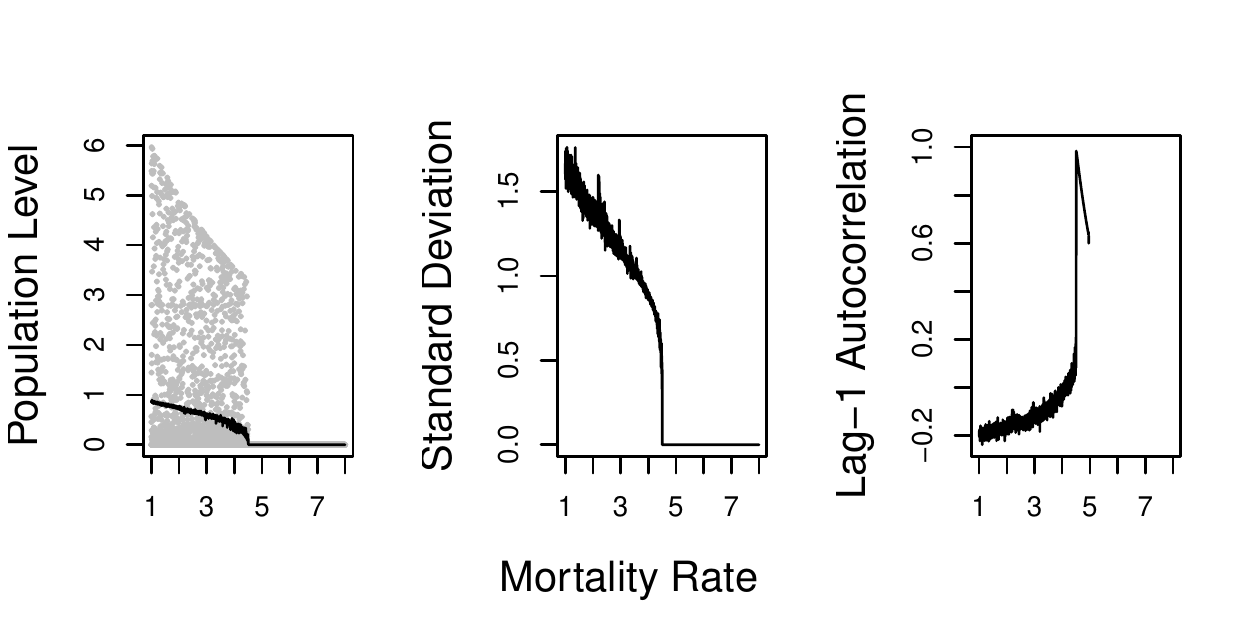}
\caption{A system where variance decreases prior to a population
collapse; adapted from Schreiber (2003). In this model, prey species
with high growth rates exhibit chaotic dynamics under predation, but
populations collapse when predation increases beyond a threshold value.
Left: The population level as a function of predation rate. Mean
dynamics shown as black line, realizations with varying initial
conditions shown as grey dots; see Schreiber (2003). Middle: Variance of
the prey population level. Note that it \emph{decreases} as predation
rate approaches the threshold. Right: Lag-1 Autocorrelation in prey
population dynamics increases as the threshold is approached}
\end{figure}

Examples are not restricted to chaotic dynamics. An example is found in
Schreiber and Rudolf (2008), in which variance is observed to decrease
before a sudden transition that results in the extinction of the
population.

Another non-chaotic example is found in some spatially extended systems
that exhibit a type of bifurcation not accompanied by CSD. In this class
of models, individual locations are subject to saddle node-type regime
shifts and influence adjacent locations via short-range facilitation and
long-range competition. Such models are used represent transitions
between vegetation types in response to changing water availability, and
reproduce naturally occurring vegetation patterns (Rietkerk and van de
Koppel 2008). In such systems, a regime shift in one location can
propagate spatially and transition the whole system from one regime to
another. Such a transition occurs if the control parameter (e.g.,
rainfall), exceeds the \emph{Maxwell point} - the value at which a local
disturbance propagates outwards (Bel et al. 2012). The Maxwell point may
be far from the level at which an individual location would undergo a
saddle-node bifurcation, and thus the system's global dynamics would not
exhibit CSD prior to such a transition. This case illustrates the
importance of distinguishing between \emph{local} and \emph{global}
system dynamics and identifying the appropriate scale of observation.

Finally, Boerlijst et al. (2013) found that indicators of CSD do not
appear prior to saddle-node bifurcations when perturbations are not in
the direction of a system's dominant eigenvalue, and even then may only
appear in one variable of the system. In their example case, increased
variance and autocorrelation only occurred when noise was applied to the
juvenile population of a model with juveniles, adults, and predators,
and it did not appear when identical noise was applied to all three.
When CSD indicators did appear, they only did so in the juvenile
population variables. This represents another under-explored area -
selecting appropriate variables for early-warning detection in
multivariate systems. Even where CSD is present, it may not be expressed
in all system components.

\subsection{Non-Catastrophic Bifurcations Preceded by CSD (III)}

Not all regime shifts are rapid. Some systems undergo bifurcations
between qualitatively different, but quantitatively similar regimes.
These transitions may be reversible. In a management setting, such
qualitative changes may be gradual, so warning signals that detect such
transitions may be effective ``false positives.''

CSD precedes several types of these non-catastrophic bifurcations. In
the subcritical form of a Hopf bifurcation, a system transitions from a
stable equilibrium to a stable cycle. As a control parameter approaches
the critical threshold, the system's dominant eigenvalue approaches zero
and thus exhibits CSD (Chisholm and Filotas 2009, Kéfi et al. 2012).
However, the mean value of the equilibrium does not change dramatically,
and the transition from stable equilibrium to cycles is gradual as the
cycle sizes grow from zero at the threshold value. To appreciate how
this bifurcation is gradual rather than catastrophic, note that in the
presence of stochasticity, the system behavior observed on either side
of the threshold may be indistinguishable: on one side stochasticity
bounces the system around a stable node, while on the other it bounces
the system around a very small limit cycle in the same region of state
space. Even when oscillations grow quickly, returning the environmental
conditions (bifurcation parameter) to the previous conditions restores
the stable node -- the bifurcation does not exhibit the hysteresis of
the saddle node bifurcation. Contrast this to a critical transition in
which any stochastic fluctuation across the threshold could lead to a
qualitatively different state.

The system's eigenvalue also passes through zero in the case of the
transcritical bifurcation. The transcritical is a degenerate case of the
saddle-node, and occurs in many of the same systems. However, when a
system passes through a transcritical bifurcation, the stable
equilibrium transitions smoothly from positive to zero, or the reverse.
In population systems, this corresponds to a transition from an
equilibrium of a very small population size to extinction - an important
but non-catastrophic, and probably directly observable, event. CSD is
observed prior to the transcritical bifurcations (Chisholm and Filotas
2009, Kéfi et al. 2012).

An experimental example of a transcritical bifurcation is found in Drake
and Griffen (2010), where a population of \emph{Daphnia} was forced
through a transcritical bifurcation by reducing food supplies and
driving population growth rates below zero. Indicators of CSD
(variation, skewness, autocorrelation, and spatial correlation)
increased prior to collapse of the population.

\subsection{CSD in the absence of bifurcations or regime shifts. (IV)}

Critical slowing down may appear in systems without any bifurcations.
Kéfi et al. (2012) showed that smooth transitions that modify a system's
potential and decrease the value of its dominant eigenvalue would result
in longer return times and greater variance and autocorrelation in
system behavior (See Figure 3). When the transition between states is
smooth, these measures will exhibit a smooth increase to a maximum and
then a decrease, unlike the sharp peaks found in systems with
bifurcations. Nonetheless, both exhibit increasing measures of CSD that
may be indistinguishable.

\begin{figure}[htbp]
\centering
\includegraphics{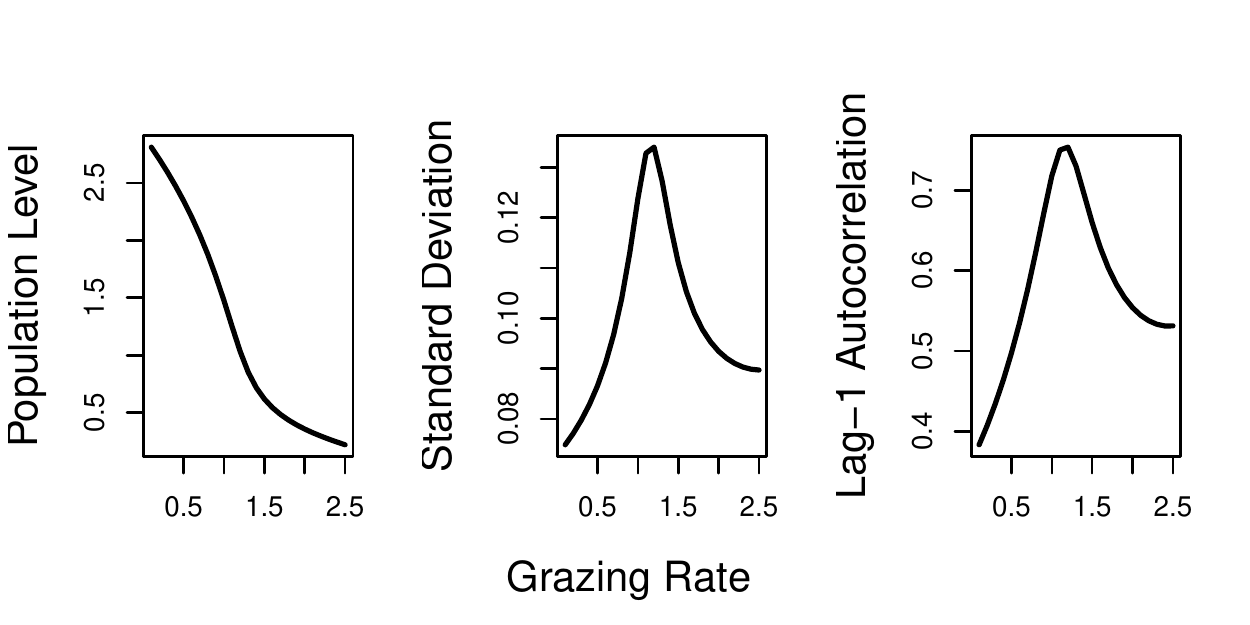}
\caption{A system where critical slowing down is observed without a
critical threshold, from Kéfi et al. (2012). In this model, prey have
logistic growth and are subject to predation with a Type III functional
response, but there is no bifurcation. Instead, average prey population
exhibits a smooth response to increased predation (grazing). Left: The
population level as a function of predation rate. Middle: Variance of
the prey population level. Right: Lag-1 Autocorrelation in prey
population dynamics as grazing rate increases. Note that both indicators
increase despite the lack of a bifurcation.}
\end{figure}

\subsection{Catastrophic Regime Shifts without Bifurcations or CSD (V)}

Some rapid regime shifts are not due to bifurcations at all. A large
external forcing (as illustrated in Figure 4) may change the behavior of
a system without any warning. This mechanism is commonly recognized,
(Scheffer et al. 2001, 2009, Barnosky et al. 2012, Scheffer et al.
2012), but others are possible. An internal stochastic event may switch
a system between dynamic regimes, or a change in system behavior may be
the manifestation of a long-term transient. In none of these cases would
CSD be expected to precede such changes. Nonetheless, it may be
difficult to distinguish such cases from bifurcations.

\begin{figure}[htbp]
\centering
\includegraphics{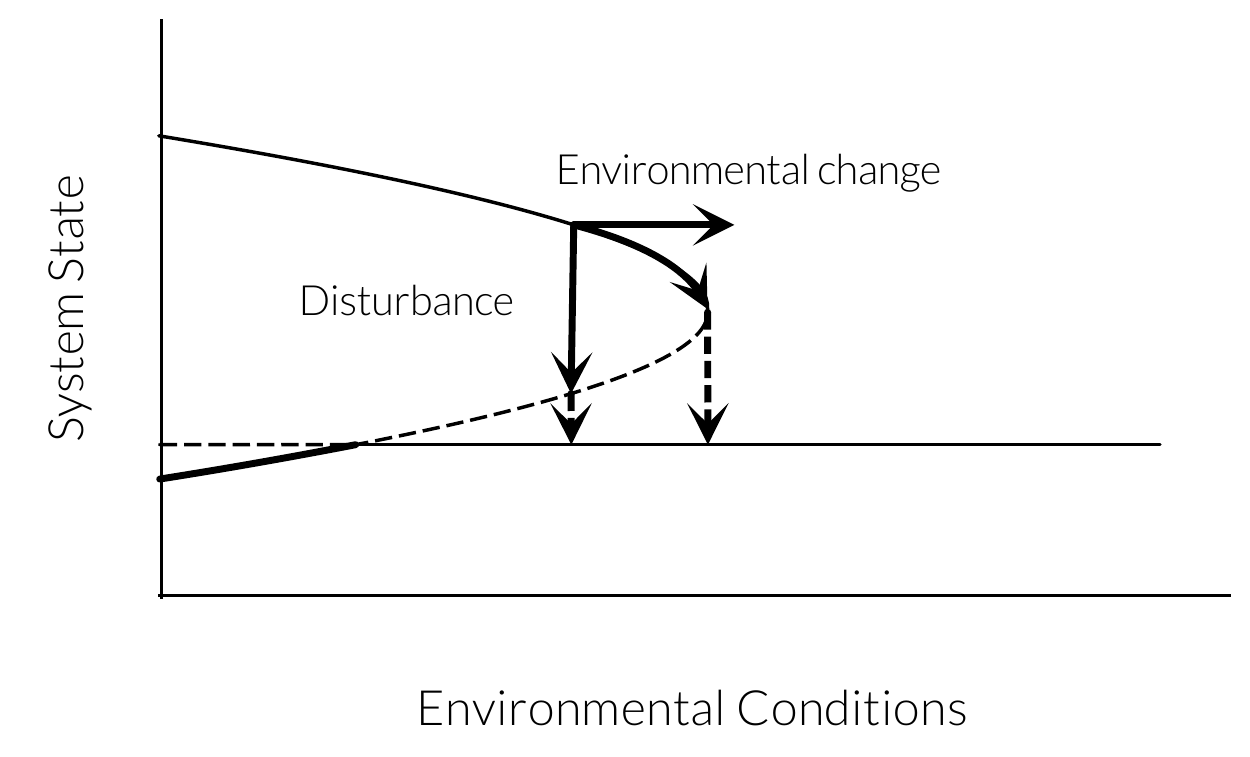}
\caption{Difference between different types of perturbations. On the
horizontal axis is the bifurcation parameter, representing the state of
the environment (e.g.~annual mean temperature) whose slow change could
lead to a sudden shift. A direct disturbance to the system state
(e.g.~population size, vertical axis) could also cause a transition if
it is large enough to cross the stability threshold (dashed line). Such
a perturbation can come from exogenous factors such as anthropogenic
pressures or occur by chance from intrinsic stochasticity. These
distinct mechanisms of disturbance and environmental change are coupled
-- as the environment deteriorates, moving the system right on the
diagram, the probability that a disturbance crosses the threshold
increases. From Bel et al. (2012).}
\end{figure}

Large, rapid changes in external conditions will result in rapid changes
in ecological system dynamics. For instance, rapid changes in North
American vegetation at the start of the Bølling-Allerød and end of the
Younger Dryas period are thought to be responses to similarly large,
rapid changes in climate (Williams et al. 2011). Doney and Sailley
(2013) interpret a recent analysis by Di Lorenzo and Ohman (2013) as
demonstrating that what were previously thought of as regime shifts in
krill dynamics in the Pacific ocean (Hare and Mantua 2000) could
actually be explained by a close coupling to the external forcing of El
Nino environmental dynamics through the Pacific Decadal Oscillation
(PDO). Schooler et al. (2011) found that lakes with the invasive plant
\emph{Salvaniai molesta} and herbivorous weevils alternated between low-
and high-\emph{Salvnia} states driven by disturbances from regular
external flooding events. These examples highlight cases that involve
critical transitions between regimes under circumstances that do not
permit the discovery of early warning signals, as CSD is not anticipated
under these mechanisms.

Internally-driven stochastic perturbations may shift systems from one
state to another even if underlying environmental conditions remain the
same. In such conditions EWS would not be expected. Hastings and Wysham
(2010) showed that in a model where one species with stochastic Ricker
dynamics disperses among eight patches, model behavior can switch
stochastically between wildly oscillatory behavior and regularly cycling
regimes even while parameters (including stochastic variability) remain
the same. Ditlevsen and Johnsen (2010) examined 25 abrupt climate
changes that occurred during the last glacial period (Dansgaard-Oeschger
events) and found no evidence for CSD in high-resolution climate data
from ice cores, and concluded that the events were driven by endogenous
climate stochasticity rather than regime shifts (though see Cimatoribus
et al. 2013 for an alternative conclusion).

Some events that appear to be regime shifts may actually be transients
in some systems. Sudden changes in dynamics can occur in simple
ecological models with strong density dependence that take long times to
reach equilibrium. Hastings (1998) showed such dynamics in model of
dispersal of inter- or sub-tidal organisms whose larvae disperse along a
coastline. Over the thousands of years it takes the model to reach
equilibrium, it may alternate between temporary regimes of regular
cycles and chaos that switch in only a few years. While on long time
scales these are technically not regime shifts, such changes would
effectively appear to be regime shifts on shorter ones. We would not
expect such regime shifts to be preceded with CSD.

Of course, stochastically-driven regime shifts may occur in systems
where bifurcations are also possible, and it may be difficult to
distinguish between the two. Renne et al. (2013), for example suggest
that ecosystems were under near-critical stress due to climate changes
just prior to the Chicxulub meteor impact, which resulted in mass
extinction. In such a case, EWS may precede the regime shift even if it
is ultimately triggered by a stochastic event.

\section{Statistical problems in detecting early warning signals}

The above cases show that behavior providing EWS before regime shifts
may only be present in certain types of ecological systems (e.g.~see the
conditions outlined in Scheffer et al. 2009). An additional important
consideration is whether these behaviors will be \emph{detectable}. To
be usable as EWS, system behavior must be detectable well enough in
advance of a regime shift to serve in decision-making, and be reliably
distinguishable from other patterns.

Ecological data is often sparse, noisy, autocorrelated and subject to
confounding driving variables, in contrast to much of the experimental
or simulated data used to test EWS. Under common levels of noise found
in field data, CSD-based EWS often fail (Perretti and Munch 2012).

A wide variety of statistical summary indicators have been examined as
potential detectors of CSD. The most common are variance and
autocorrelation. Others include skewness (Guttal and Jayaprakash 2008a)
and conditional heteroscedasticity (Seekell et al. 2011). These
statistics are typically calculated on sliding windows of time-series
data and tested formally or informally for trends. The relative power of
these tests varies considerably with context; no indicator has
consistently outperformed others (Dakos et al. 2011b, 2012, Lindegren et
al. 2012, Perretti and Munch 2012). Also, measuring these indicators
requires making sometimes arbitrary calculations. For instance, the
power of lag-1 autocorrelation to detect a regime shift may be modified
by changing methods of data aggregation, de-trending, changing sliding
window length, filtering signal bandwidth (Lenton et al. 2012). These
choices may be optimized when enough calibration data is available, as
Lenton et al. (2012) were able to do with several sets of paleoclimate
data. However, such calibration may not be possible with many ecological
datasets. Multiple-method (Lindegren et al. 2012) and composite indices
(Drake and Griffen 2010) have been proposed, but their power relative to
other indicators is unknown.

Another approach to detecting CSD has been fitting time series data to
models. Two approaches have been used for these model-based methods.
First, models may be used to calculate summary statistics related to
CSD, such as eigenvalues (Lade and Gross 2012) or diffusion terms in
jump-diffusion models (Carpenter 2011, Brock and Carpenter 2012). These
statistics are then examined for trends in the same fashion as the
summary statistics above. Alternatively, models representing both
deteriorating and stable conditions may be fit to the data and in order
to determine which is more likely (Dakos et al. 2012, ). Boettiger and
Hastings (2012a) found that likelihood ratio tests were more powerful
than trend-based summary statistic tests across several real and
simulated ecological data sets. This approach is also more robust than
summary-statistic methods to spurious correlations that arise when
collapses are driven by purely stochastic events (Boettiger and Hastings
2012a).

Care is required in the criteria used to judge the power of warning
signal methods. The trade-off between false negatives and false
positives is a matter of not just statistical but economic efficiency.
For instance, a large number of false positives may be acceptable if
they reduce the probability of a false warning that would result in an
otherwise avoidable catastrophic regime shift, and the costs of failing
to detect such a shift exceed that of the false positives. Boettiger and
Hastings (2012a) suggest the use of receiver-operating characteristic
(ROC) curves to describe the performance of various EWS. ROC curves
(Figure 5) represent the false positive rate at any true positive rate.
The area under the curve (AUC) is a useful metric of overall
performance. AUC will be one if the signal is perfect and 0.5 if the
signal performs no better than random. The complete shape of the curve
provides more information on the possible trade-offs under different
sensitivities. This information, combined with a decision-theoretic
framework, has the potential to illuminate the cases in which EWS can be
useful.

\begin{figure}[htbp]
\centering
\includegraphics{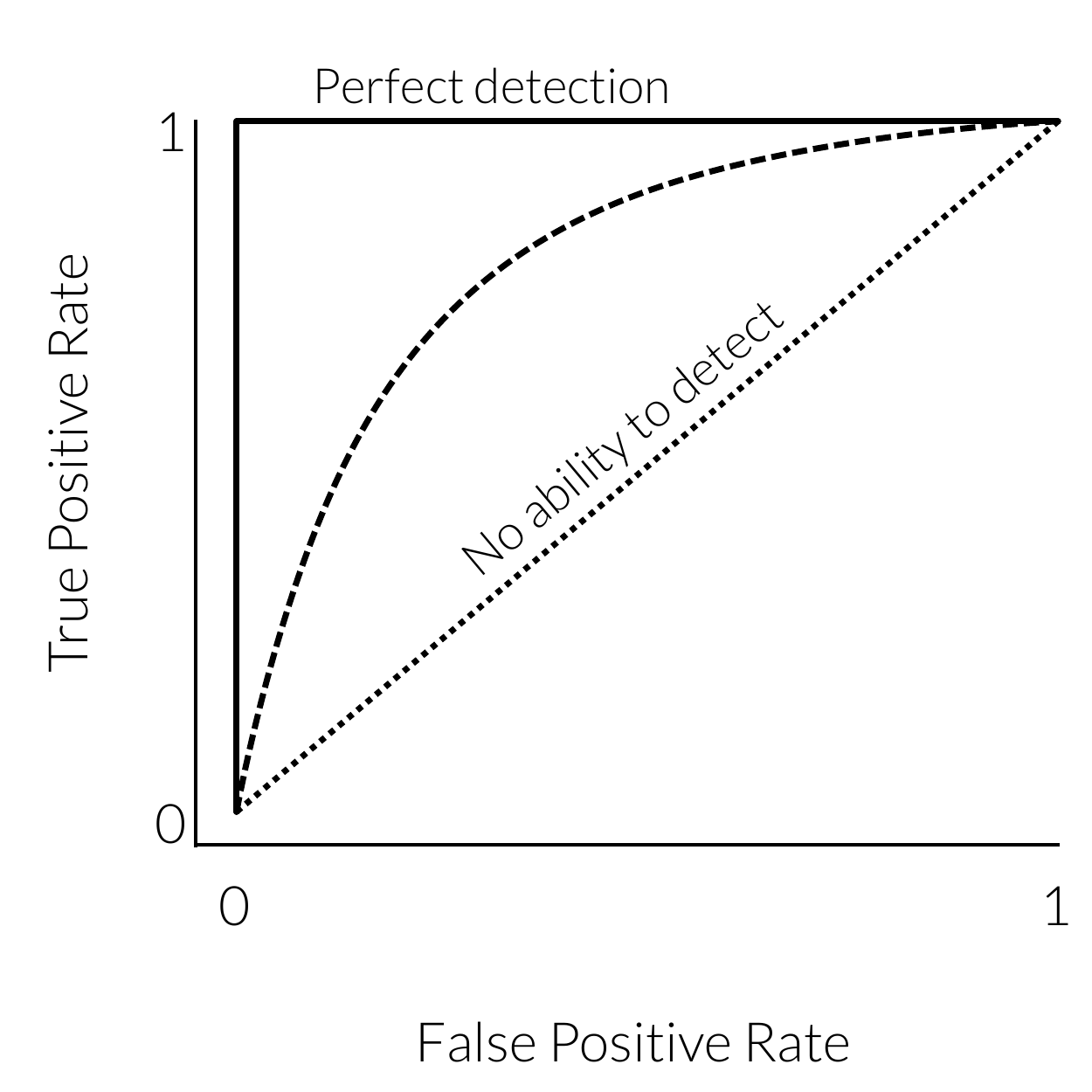}
\caption{Receiver-operating characteristic (ROC) curves illustrate the
trade-off between false positive and true positive detection rates of an
EWS. Perfect warning signals (solid curve) would identify all thresholds
while generating no false positives, while very poor signals would have
no ability to distinguish false from true signals (dotted line). In
reality, warning signals' have a trade-off between the two which is
described by a curve (dotted line) or summarized by the area under the
ROC curve}
\end{figure}

\section{Discussion}

Recognizing the potential for early warning signals of critical
transitions represents a substantial leap forward in addressing one of
the most challenging questions in ecology and ecosystem management
today. In the decades prior, the prospect that ecosystems could make
sudden transitions into an undesirable state due to gradual, slow
changes in their environment hung like a specter over both our
understanding and management of natural systems. Research that points to
the possibility of detecting these transitions holds the promise of
meeting this challenge and has attracted justifiably widespread
attention among both theoretical and empirical communities. Nonetheless,
our understanding of early warning signals is still in its infancy. Thus
far, our best understanding and empirical experience lies in transitions
that are driven by saddle-node bifurcations.

While saddle-node bifurcations may be common, they represent only part
of the potential mechanisms for rapid regime shift. Occupying the center
of our diagram, Figure 1, such transitions represent our best-understood
cases. Researchers have relied on existing expertise and prior research
to identify empirical systems most likely to experience critical
transitions through the saddle-node-like mechanism (e.g. Carpenter et
al. 2011, Dai et al. 2012), and have achieved a close match to
theoretical predictions of early warning signals. While these examples
provide a much needed proof-of-principle that these signals can be
detected in the real world, it is too early to apply the same methods to
novel systems where the saddle-node is only one of many possible
mechanisms. We are not yet able to determine if a natural system is
likely to have a saddle-node bifurcation without detailed study, despite
the popularity of saddle-node models.

Thus, establishing the saddle node mechanism is a necessary condition of
using CSD as a warning signal. This can be done via manipulation in
simple experimental systems (Veraart et al. 2012, Dai et al. 2012), but
this is impractical in most natural systems. Another approach is to
assume the saddle-node mechanism applies to a limited set of systems
that have well-studied examples, such as lakes undergoing eutrophication
(Scheffer et al. 2001), lakes with `trophic-triangle' cascade mechanisms
(Carpenter and Kitchell 1996, Walters and Kitchell 2001, Carpenter et
al. 2008), forest/savannah transitions (Staver et al. 2011, Hirota et
al. 2011), and rangeland transitions (Walker1993; Anderies et al. 2002).
Fitting simplified saddle-node models to past regime shifts (Boettiger
and Hastings 2012a) in less well-understood systems may provide evidence
for the mechanism. However, care must be taken to specify sufficient
alternative models.

CSD alone cannot be used as evidence of regime shifts. In some cases, it
will be present when no transition is approaching. In other cases,
regime shifts occur without CSD. Though false alarms and missed events
can occur in any statistical procedure, the cases discussed here
demonstrate that these errors will also arise when the underlying
dynamics do not correspond to our assumptions. These situations fall in
the uncharted area beyond the center of Figure 1, where research has
just begun to illuminate their existence and properties. A better
theoretical and empirical understanding of these cases will allow us to
construct novel warning signals, that may be opposite the patterns
observed in the familiar saddle-node bifurcations. Before early warning
signals can be applied in novel systems, additional information is
needed in order to determine the best signal to use.

One area that requires further exploration is the effect of different
forms of stochasticity on the existence of EWS and signal detectability.
Many processes contribute to stochastic behavior in ecological systems,
and different forms of stochasticity have different effects on system
behavior far beyond greater variance (Melbourne and Hastings 2008).
Hastings and Wysham (2010) argued that most examples of detectable CSD
indicators were found in models with additive stochasticity and smooth
potentials. Boerlijst et al. (2013), however, found that stochasticity
had the same effects whether it was additive or included in the
population growth rate. Instead, they found the \emph{direction} of
stochastic perturbations relative to the system's eigenvalue determined
whether CSD indicators were detectable. The form of stochasticity may be
important in the detectability of CSD indicators even where CSD is
present, because stochastic perturbations are needed to explore system
state-space, while at the same time can reduce the statistical power.
More work such as Perretti and Munch (2012), which examined the role of
noise color in detecting CSD, will be useful.

Another area that has is understanding how the relationship between the
scale of observation and the scale of ecological processes affect the
efficacy of EWS. As shown by the Maxwell Point example in Bel et al.
(2012), EWS which detect local bifurcations may not detect global
bifurcations in system behavior. The scale of observation likely also
will affect the statistical power of EWS. Similarly, as illustrated in
Boerlijst et al. (2013), the choice of variables to observe in
multivariate systems is important, but little is known about how to
select the appropriate variable for detecting EWS.

The future of early warning signals lies in the uncharted territory. For
certain classes of transitions, such as stochastically-driven regime
shifts, prediction may not be possible. In such cases, management
options include optimizing outcomes despite the possibility of regime
shifts, or possibly taking actions to reduce the long-term probability
of regime shifts, despite short-term unpredictability. Likewise, regime
shifts driven by external perturbation or strong forcing are not
predictable \emph{if} the scope of management does not include the
external causes. Proper scoping of the management problem can avoid this
situation (Fischer et al. 2009, Alliance 2010, Polasky et al. 2011).
More research is needed in methods of distinguishing such cases from
those in which early detection may be possible.

For other classes of transitions, prediction may be possible but other
EWS must be explored. Flickering (Brock and Carpenter 2010, Wang et al.
2012), or rapid transitions between states prior to a more permanent
transition, is one signal that may apply across many types of systems.
It manifests in bi-modality and high variance in times series. Spatial
pattern development may be a warning signal in systems with
short-distance positive feedbacks but long-distance negative feedbacks,
such as grassland-desert transitions (Rietkerk et al. 2004). Other
spatial signals may apply where systems include both saddle nodes and
positive feedbacks across space (Litzow et al. 2008, Guttal and
Jayaprakash 2008a, Dakos et al. 2009, Bailey 2010, Dakos et al. 2011b,
Bel et al. 2012). A critical task for EWS research is to map these
signals to their domains of applicability, and create methods to
establish if ecosystems fall into these domains.

\section{Acknowledgments}

This work was partially supported by the Center for Stock Assessment
Research, a partnership between the University of California Santa Cruz
and the Fisheries Ecology Division, Southwest Fisheries Science Center,
Santa Cruz, CA, to CB; the NSF Integrative Graduate Education and
Research Traineeship Program to NR and by funding from NSF Grant EF
0742674 to AH.

\section{References}

Alliance, R. 2010. Assessing Resilience in Social-Ecological Systems :
Workbook for Practitioners. Version 2.0.

Anderies, J. M., M. A. Janssen, and B. H. Walker. 2002. Grazing
Management, Resilience, and the Dynamics of a Fire-driven Rangeland
System. Ecosystems 5:23--44.

Bailey, R. M. 2010. Spatial and temporal signatures of fragility and
threshold proximity in modelled semi-arid vegetation.. Proceedings.
Biological sciences / The Royal Society.

Barnosky, A. D., E. A. Hadly, J. Bascompte, E. L. Berlow, J. H. Brown,
M. Fortelius, W. M. Getz, J. Harte, A. Hastings, P. A. Marquet, N. D.
Martinez, A. Mooers, P. Roopnarine, G. Vermeij, J. W. Williams, R.
Gillespie, J. Kitzes, C. Marshall, N. Matzke, D. P. Mindell, E. Revilla,
and A. B. Smith. 2012. Approaching a state shift in Earth's biosphere.
Nature 486:52--58.

Bel, G., A. Hagberg, and E. Meron. 2012. Gradual regime shifts in
spatially extended ecosystems. Theoretical Ecology 5:591--604.

Boerlijst, M. C., T. Oudman, and A. M. de Roos. 2013. Catastrophic
Collapse Can Occur without Early Warning: Examples of Silent
Catastrophes in Structured Ecological Models. PLoS ONE 8:62033.

Boettiger, C., and A. Hastings. 2012a. Early warning signals and the
prosecutor's fallacy. Proceedings of the Royal Society B: Biological
Sciences 279:4734--4739.

Boettiger, C., and A. Hastings. 2012a. Quantifying limits to detection
of early warning for critical transitions.. Journal of the Royal
Society, Interface / the Royal Society 9:2527--39.

Brock, W. A., and S. R. Carpenter. 2010. Interacting Regime Shifts in
Ecosystems: Implication for Early Warnings. Ecological Monographs
80:353--367.

Brock, W. A., and S. R. Carpenter. 2012. Early Warnings of Regime Shift
When the Ecosystem Structure Is Unknown. PLoS ONE 7:45586.

Carpenter, S. R. 2011. Early warnings of unknown nonlinear shifts : a
nonparametric approach. Ecology 92:2196--2201.

Carpenter, S. R., W. A. Brock, J. J. Cole, J. F. Kitchell, and M. L.
Pace. 2008. Leading indicators of trophic cascades.. Ecology letters
11:128--38.

Carpenter, S. R., J. J. Cole, M. L. Pace, R. Batt, W. A. Brock, T.
Cline, J. Coloso, J. R. Hodgson, J. F. Kitchell, D. a Seekell, L. Smith,
and B. Weidel. 2011. Early Warnings of Regime Shifts: A Whole-Ecosystem
Experiment. Science 1079.

Carpenter, S. R., D. Ludwig, and W. A. Brock. 1999. Management of
eutrophication for lakes subject to potentially irreversible change.
Ecological Applications 9:751--771.

Carpenter, S. R., and W. A. Brock. 2006. Rising variance: a leading
indicator of ecological transition.. Ecology letters 9:311--8.

Carpenter, S. R., and W. a Brock. 2010. Early warnings of regime shifts
in spatial dynamics using the discrete Fourier transform. Ecosphere
1:10.

Carpenter, S. R., and J. F. Kitchell. 1996. The trophic cascade in
lakes. . Cambridge University Press.

Chisholm, R. A., and E. Filotas. 2009. Critical slowing down as an
indicator of transitions in two-species models.. Journal of theoretical
biology 257:142--9.

Cimatoribus, S. S. Drijfhout, V. Livina, and G. van der Schrier. 2013.
Dansgaard--Oeschger events: bifurcation points in the climate system.
Climate of the Past 9:323--333.

Costantino, R. F., R. A. Desharnais, J. M. Cushing, and B. Dennis. 1997.
Chaotic Dynamics in an Insect Population. Science 275:389--391.

Dai, L., D. Vorselen, K. S. Korolev, and J. Gore. 2012. Generic
Indicators for Loss of Resilience Before a Tipping Point Leading to
Population Collapse. Science 336:1175--1177.

Dakos, V., S. R. Carpenter, W. a Brock, A. M. Ellison, V. Guttal, A. R.
Ives, S. Kéfi, V. Livina, D. a Seekell, E. H. van Nes, and M. Scheffer.
2012. Methods for Detecting Early Warnings of Critical Transitions in
Time Series Illustrated Using Simulated Ecological Data. PLoS ONE
7:41010.

Dakos, V., S. Kéfi, M. Rietkerk, E. H. van Nes, and M. Scheffer. 2011a.
Slowing down in spatially patterned ecosystems at the brink of
collapse.. The American naturalist 177:153.

Dakos, V., E. H. Nes, R. Donangelo, H. Fort, and M. Scheffer. 2009.
Spatial correlation as leading indicator of catastrophic shifts.
Theoretical Ecology:163--174.

Dakos, V., E. H. van Nes, P. D'Odorico, and M. Scheffer. 2011b.
Robustness of variance and autocorrelation as indicators of critical
slowing down. Ecology.

Dakos, V., M. Scheffer, E. H. van Nes, V. Brovkin, V. Petoukhov, and H.
Held. 2008. Slowing down as an early warning signal for abrupt climate
change.. Proceedings of the National Academy of Sciences 105:14308--12.

Di Lorenzo, E., and M. D. Ohman. 2013. A double-integration hypothesis
to explain ocean ecosystem response to climate forcing.. Proceedings of
the National Academy of Sciences of the United States of America.

Ditlevsen, P. D., and S. J. Johnsen. 2010. Tipping points: Early warning
and wishful thinking. Geophysical Research Letters 37:2--5.

Doney, S. C., and S. F. Sailley. 2013. When an ecological regime shift
is really just stochastic noise.. Proceedings of the National Academy of
Sciences of the United States of America 110:2438--2439.

Drake, J. M., and B. D. Griffen. 2010. Early warning signals of
extinction in deteriorating environments. Nature 467:456--7.

Fischer, J., G. D. Peterson, T. A. Gardner, L. J. Gordon, I. Fazey, T.
Elmqvist, A. Felton, C. Folke, and S. Dovers. 2009. Integrating
resilience thinking and optimisation for conservation.. Trends in
ecology \& evolution 24:549--554.

Gandhi, A., S. Levin, and S. Orszag. 1998. ``Critical slowing down'' in
time-to-extinction: an example of critical phenomena in ecology. Journal
of theoretical biology 192:363--76.

Grebogi, C., E. Ott, and J. A. Yorke. 1983. Crises, sudden changes in
chaotic attractors, and transient chaos. Physica D: Nonlinear Phenomena.

Guttal, V., and C. Jayaprakash. 2008a. Spatial variance and spatial
skewness: leading indicators of regime shifts in spatial ecological
systems. Theoretical Ecology 2:3--12.

Guttal, V., and C. Jayaprakash. 2008a. Changing skewness: an early
warning signal of regime shifts in ecosystems.. Ecology letters
11:450--60.

Hare, S. R., and N. J. Mantua. 2000. Empirical evidence for North
Pacific regime shifts in 1977 and 1989. Progress in oceanography
47:103--145.

Hastings, A. 1998. Transients in spatial ecological models. Modeling
spatiotemporal dynamics in ecology:189--198.

Hastings, A., C. L. Hom, S. Ellner, P. Turchin, and H. C. J. Godfray.
1993. Chaos in ecology: is mother nature a strange attractor?. Annual
Review of Ecology and Systematics 24:1--33.

Hastings, A., and D. B. Wysham. 2010. Regime shifts in ecological
systems can occur with no warning.. Ecology letters 13:464--72.

Hirota, M., M. Holmgren, E. H. Van Nes, and M. Scheffer. 2011. Global
resilience of tropical forest and savanna to critical transitions..
Science 334:232--5.

Holling, C. S. S. 1973. Resilience and Stability of Ecological Systems.
Annual Review of Ecology and Systematics 4:1--23.

Jackson, J. B., M. X. Kirby, W. H. Berger, K. a Bjorndal, L. W.
Botsford, B. J. Bourque, R. H. Bradbury, R. Cooke, J. Erlandson, J. a
Estes, T. P. Hughes, S. Kidwell, C. B. Lange, H. S. Lenihan, J. M.
Pandolfi, C. H. Peterson, R. S. Steneck, M. J. Tegner, and R. R. Warner.
2001. Historical overfishing and the recent collapse of coastal
ecosystems.. Science (New York, N.Y.) 293:629--37.

Kéfi, S., V. Dakos, M. Scheffer, E. H. van Nes, and M. Rietkerk. 2012.
Early warning signals also precede non-catastrophic transitions.
Oikos:1--8.

Kéfi, S., M. Rietkerk, C. L. Alados, Y. Pueyo, V. P. Papanastasis, A.
Elaich, P. C. de Ruiter, L. Alados, and P. C. D. Ruiter. 2007. Spatial
vegetation patterns and imminent desertification in Mediterranean arid
ecosystems.. Nature 449:213--7.

Lade, S. J., and T. Gross. 2012. Early Warning Signals for Critical
Transitions: A Generalized Modeling Approach. PLoS Computational Biology
8:1002360.

Lenton, T. M., V. N. Livina, V. Dakos, E. H. van Nes, and M. Scheffer.
2012. Early warning of climate tipping points from critical slowing
down: comparing methods to improve robustness.. Philosophical
transactions. Series A, Mathematical, physical, and engineering sciences
370:1185--204.

Lenton, T. M., R. J. Myerscough, R. Marsh, V. N. Livina, A. R. Price, S.
J. Cox, and G. Team. 2009. Using GENIE to study a tipping point in the
climate system.. Philosophical transactions. Series A, Mathematical,
physical, and engineering sciences 367:871--84.

Lewontin, R. C. 1969. The meaning of stability. Pages 13--25 \emph{in}
G. W. Woodwell and H. H. Smith, editors. Brookhaven Symposia in Biology.
. Brookhaven National Laboratory, Upton, N.Y.

Lindegren, M., V. Dakos, J. P. Gröger, A. G\textbackslash{}aa rdmark, G.
Kornilovs, S. A. Otto, and C. Möllmann. 2012. Early Detection of
Ecosystem Regime Shifts: A Multiple Method Evaluation for Management
Application. PLoS ONE 7:38410.

Litzow, M. A., J. D. Urban, and B. J. Laurel. 2008. Increased spatial
variance accompanies reorganization of two continental shelf ecosystems.
Ecological Applications 18:1331--1337.

Ludwig, D., D. D. Jones, and C. S. Holling. 1978. Qualitative Analysis
of Insect Outbreak Systems: The Spruce Budworm and Forest. The Journal
of Animal Ecology 47:315.

May, R. M. 1977. Thresholds and breakpoints in ecosystems with a
multiplicity of stable states. Nature 269:471--477.

Melbourne, B. A., and A. Hastings. 2008. Extinction risk depends
strongly on factors contributing to stochasticity.. Nature 454:100--3.

Mumby, P. J., A. Hastings, and H. J. Edwards. 2007. Thresholds and the
resilience of Caribbean coral reefs.. Nature 450:98--101.

Perretti, C. T., and S. B. Munch. 2012. Regime shift indicators fail
under noise levels commonly observed in ecological systems. Ecological
Applications 22:1772--1779.

Polasky, S., S. R. Carpenter, C. Folke, and B. Keeler. 2011.
Decision-making under great uncertainty: environmental management in an
era of global change.. Trends in ecology \& evolution 26:398--404.

Renne, P. R., A. L. Deino, F. J. Hilgen, K. F. Kuiper, D. F. Mark, W. S.
Mitchell, L. E. Morgan, R. Mundil, and J. Smit. 2013. Time scales of
critical events around the Cretaceous-Paleogene boundary.. Science
339:684--7.

Rietkerk, M., J. van de Koppel, S. C. Dekker, and P. C. de Ruiter. 2004.
Self-organized patchiness and catastrophic shifts in ecosystems..
Science 305:1926--9.

Rietkerk, M., and J. van de Koppel. 2008. Regular pattern formation in
real ecosystems.. Trends in ecology \& evolution 23:169--75.

Scheffer, M., J. Bascompte, W. A. Brock, V. Brovkin, S. R. Carpenter, V.
Dakos, H. Held, E. H. van Nes, M. Rietkerk, and G. Sugihara. 2009.
Early-warning signals for critical transitions.. Nature 461:53--9.

Scheffer, M., S. R. Carpenter, J. A. Foley, C. Folke, and B. Walker.
2001. Catastrophic shifts in ecosystems.. Nature 413:591--6.

Scheffer, M., S. R. Carpenter, T. M. Lenton, J. Bascompte, W. A. Brock,
V. Dakos, J. van de Koppel, I. A. van de Leemput, S. A. Levin, E. H. van
Nes, M. Pascual, and J. Vandermeer. 2012. Anticipating Critical
Transitions. Science 338:344--348.

Schooler, S. S., B. Salau, M. H. Julien, and A. R. Ives. 2011.
Alternative stable states explain unpredictable biological control of
Salvinia molesta in Kakadu. Nature 470:86--89.

Schreiber, S. J. 2003. Allee effects, extinctions, and chaotic
transients in simple population models. Theoretical Population Biology
64:201--209.

Schreiber, S. J., and S. Rittenhouse. 2004. From simple rules to cycling
in community assembly. Oikos 93:430--358.

Schreiber, S., and V. H. W. Rudolf. 2008. Crossing habitat boundaries:
coupling dynamics of ecosystems through complex life cycles.. Ecology
letters 11:576--87.

Seekell, D. A., S. R. Carpenter, and M. L. Pace. 2011. Conditional
heteroscedasticity as a leading indicator of ecological regime shifts..
The American naturalist 178:442--51.

Staver, A. C., S. Archibald, and S. Levin. 2011. Tree cover in
sub-Saharan Africa: rainfall and fire constrain forest and savanna as
alternative stable states.. Ecology 92:1063--72.

Veraart, A. J., E. J. Faassen, V. Dakos, E. H. van Nes, M. Lürling, and
M. Scheffer. 2012. Recovery rates reflect distance to a tipping point in
a living system.. Nature 481:357--9.

Walters, C., and J. F. Kitchell. 2001. Cultivation/depensation effects
on juvenile survival and recruitment: implications for the theory of
fishing. Canadian Journal of Fisheries and Aquatic Sciences 58:39--50.

Wang, R., J. a Dearing, P. G. Langdon, E. Zhang, X. Yang, V. Dakos, and
M. Scheffer. 2012. Flickering gives early warning signals of a critical
transition to a eutrophic lake state.. Nature:1--4.

Williams, J. W., J. L. Blois, and B. N. Shuman. 2011. Extrinsic and
intrinsic forcing of abrupt ecological change: case studies from the
late Quaternary. Journal of Ecology 99:664--677.

Wissel, C. 1984. A universal law of the characteristic return time near
thresholds. Oecologia 65:101--107.

\end{document}